\documentstyle[preprint,aps]{revtex}
\draft
\begin{document}

\title{Near-threshold $K^{+}$ production in Heavy-ion Collisions}
\bigskip
\author{\bf Bao-An Li}
\address{Cyclotron Institute and Department of Physics\\
Texas A\&M University, College Station, TX 77843}
\maketitle

\begin{abstract}
Within a hadronic transport model we study in detail contributions to
kaon yields and momentum spectra from various baryon (resonance)-baryon
(resonance) and $\pi N$ interactions in heavy-ion collisions at beam energies
near the free-space kaon production threshold. It is found that
the finite lifetime of baryon resonances affects significantly the shape of
kaon spectra, and the high energy parts of the kaon spectra are dominated by
kaons from $\pi N\rightarrow \Lambda K^{+}$ processes. $N^{*}(1440)$ resonances
are found to contribute about 10\% to the
kaon yield. Effects of boosting the Fermi momentum
distributions of the two colliding nuclei into their center of mass frame,
centrality of the reaction as well as the nuclear equation of state on
kaon yields and spectra are also discussed.
Model calculations on $K^{+}$, $\pi^{+}$ and $\pi^{-}$ spectra for the
reaction of Au+Au at $E_{beam}/A= 1.0$ GeV are compared with the
experimental data from the KaoS collaboration.
\end{abstract}

\section{Motivation}
Kaons have long been proposed as one
of the most promising messengers of the primary violent stage
of relativistic heavy-ion collisions \cite{randrup}.
Based on nuclear transport model calculations,
it was further demonstrated that kaon production in heavy-ion collisions at
subthreshold energies may be very useful in pinning down
the nuclear equation of state \cite{aichelin85}. Moreover, it has been
shown recently that kaons are also very useful in studying in-medium
properties of hadrons in the hot and dense zone formed in the
reaction \cite{feng}. However, due to the complexity of the reaction
dynamics and uncertainties in elementary kaon production processes,
the physics extracted from kaon data is still very
limited despite extensive efforts in the past years.
For a recent review of the experimental
and theoretical study on kaons, we refer the reader to
refs. \cite{grosse,metag}.

The most recent data on kaon production from the reaction of Au+Au
at a beam energy of 1.0 GeV/nucleon taken by the KaoS
collaboration \cite{senger} at SIS/GSI has further stimulated
much interest and theoretical work on the mechanism of kaon production,
the extraction of the nuclear equation of state and
in-medium properties of kaons. Most of these studies are based on
transport models\cite{lang,ko} or quantum molecular dynamics
\cite{huang,aichelin}. However, due to large discrepancies among
the results of different model calculations, the interpretation
of the experimental data has been difficult. To better understand
the discrepancies among previous
model calculations and most importantly to help interpretating the
experimental data more accuruately, we perform a study on several
aspects of kaon production using another independent transport model for
relativistic heavy-ion collisions detailed in refs. \cite{li91a,li93}.
We claim, by no means, that the model used here is superior to any
model used previously in studying kaon production. But the aspects
discussed in the following are useful for improving some of the previous
calculations.

Firstly, in all of the previous calculations, one either uses
the frozen-resonance approximation (assume resonances have infinite lifetime)
or let the resonances decay but assume that kaons from $\pi N$ collisions
can be negelected on the ground that $\pi N$ collisions only contributate
about 25\% to the total kaon yields\cite{cugnon,batko90,xiong90}.
It was then concluded that the kaon yield in heavy-ion collisions
at beam energies around 1.0 GeV/nucleon is dominated by $N\Delta$,
$\Delta\Delta$ and $NN$ collisions sequentially. However, it has been shown by
several authors that the production dynamics, multiplicity and
spectra of pions cannot be properly described by using the frozen-resonance
approximation\cite{li91a,gale,wolf,danielewicz}. Since kaons calculated
perturbatively in all of the models are almost completely determined by the
$\pi,N,\Delta,N^{*}$ dynamics, it is therefore worth studying
how the conclusions reached earlier may be affected by the finite lifetime of
baryon resonances. Indeed, we found that although $\pi N$ collisions only
contributate about 25\% to the total kaon yield, it is comparable to
contributions of most individual baryon-baryon collisions and should
not be negelected. The contribution from $\pi N$ collisions actually
dominates the high energy part of kaon spectra. The shapes of kaon
spectra calculated using infinite or finite lifetimes for baryon
resonances differ significantly.

Secondly, at beam energies around 1.0 GeV/nucelon $N^{*}(1440)$ resonances
can also be excited. Effects of the $N^{*}$ resonance on kaon yields and
spectra have not been studied. We therefore also included in the present
study contributions to kaons from the $N^{*}$ involved collisions. It is found
that the $N^{*}$ involved collisions only contribute about 10\% to the total
kaon yield and its effect on kaon spectra is minor.

Thirdly, in view of the fact that some of the previous studies
(e.g. ref.\ \cite{huang}) used the non-relativistic transformation to
boost the Fermi momentum distributions of the two colliding nuclei into their
center of mass frame, we compare kaon yields and spectra calculated
using the relativistic and the non-relativistic momentum transformation.
This seemingly numerical treatment has vitally important physical consequence.
Not only the relative contributions from $NN$ and other collision channels
differ dramatically, the total kaon yield and spectrum change as large as
that caused by changing from the soft nuclear equation of state
(with an incompressibility $K$= 200 MeV) to the stiff one ($K$= 380 MeV).

Finally, since the dynamics of kaon production is virtually determined by
the $\pi,N,\Delta,N^{*}$ dynamics, especially in the perturbative
method used in calculating kaons, and it is well known that pion
observables provide strong constraints on the reaction dynamics,
we study also whether pions from the same reaction can be
well described simultaneously as kaons by comparing $K^{+}, \pi^{+}$ and
$\pi^{-}$ spectra with the experimental data from the
KaoS collaboration.

The paper is organized as follows. In the next section
necessary inputs for calculating kaons are presented.
Our results will be presented and discussed
in section 3. A summary will be given at the end.

\section{The model and inputs}
In the present study we use the hadronic transport
model for relativistic heavy-ion collisions \cite{li91a,li93}.
It is based on the numerical solution
of a coupled set of transport equations for the phase space distribution
functions of nucleons, baryon resonances ($\Delta(1232)$ and $N^{*}(1440)$)
and pions \cite{wang}. The model has been rather successful in
studying several aspects of relativistic heavy-ion collisions.

As for the production of kaons, we use the perturbative method first
used by Randrup and Ko in ref.\ \cite{randrup} and nowadays widely used
by others. We include kaon production channels due to both
baryon-baryon collisions ( i.e. $B_1B_2\rightarrow BYK^+$,
here $B$ represents a baryon and $Y$ represents a hyperon.) as well as
pion-nucleon collisions (i.e. $\pi N\rightarrow \Lambda K^+$). More
specifically, we include kaons from the interaction of $NN, N\Delta,
\Delta\Delta, NN^{*}, \Delta N^{*}, N^{*}N^{*}$ and $\pi N$.

For kaons from baryon-baryon collisions, we use the standard
Randrup-Ko parameterizations for both the total and the differential
kaon production cross section \cite{randrup}.
For completeness we copy these parameterizations in the following.
The kaon production cross section from a nucleon-nucleon interaction
was parameterized as
\begin{equation}
\sigma _{NN\rightarrow BYK^+} (\sqrt s)=36 ~{p_{\rm max}\over m_K} ~\mu {\rm
b},
\end{equation}
where the maximum kaon momentum $p_{\rm max}$ is related to the nucleon-nucleon
center-of-mass energy $\sqrt s$ by
\begin{equation}
p_{\rm max}={1\over 2}\sqrt {\big[ s-(m_B+m_Y+m_K)^2\big]
\big[ s-(m_B+m_Y-m_K)^2\big] /s}.
\end{equation}

Kaon production cross sections from a
nucleon-delta and a delta-delta interaction were
parameterized respectively as
\begin{equation}
\sigma _{N\Delta\rightarrow BYK^+}(\sqrt s)\approx {3\over 4}
\sigma _{NN\rightarrow BYK^+}(\sqrt s),
\end{equation}
\begin{equation}
\sigma _{\Delta\Delta\rightarrow BYK^+}(\sqrt s)\approx {1\over 2}
\sigma _{NN\rightarrow BYK^+}(\sqrt s).
\end{equation}
For the $N^{*}$ involved collisions we assume that the kaon production
cross sections are the same as that in the $\Delta$ involved collisions.

In addition to the total kaon production cross section, the kaon
momentum distribution from the baryon-baryon interaction was also
parameterized by Randrup and Ko according to the phase space argument,
\begin{equation}\label{nn}
{E\over p^2}{d^3\sigma (\sqrt s)\over dpd\Omega}=
\sigma_{K^+}(\sqrt{s}){E\over 4\pi p^2}{12\over p_{\rm max}}
\left (1-{p\over p_{\rm max}}\right )\left ({p\over p_{\rm max}}\right )^2,
\end{equation}
where $\sigma _{K^+}$ is the total kaon production cross
section and $p_{\rm max}$ is the maximum kaon momentum.

For kaons from pion-nucleon collisions, we adopt the parameterization by
Cugnon \cite{cugnon}. The total kaon production cross section was
parameterized by averaging over the processes $\pi^{0} p\rightarrow
K^{+}\Lambda$ and $\pi^{+} n\rightarrow K^{+}\Lambda$ and can reproduce quite
accurately the available data up to $\sqrt{s}=3.0$ GeV.
The cross section reads
\begin{equation}
\sigma_{\pi N\rightarrow K^{+}\Lambda}(\sqrt{s})=
2.47\cdot (\sqrt{s}-\sqrt{s_{0}})~mb
\end{equation}
for $\sqrt{s_{0}}=m_{\Lambda}+m_{K}<\sqrt{s}<1.7$ GeV,
\begin{equation}
\sigma_{\pi N\rightarrow K^{+}\Lambda}(\sqrt{s})
=\frac{0.0225}{\sqrt{s}-1.6}~mb
\end{equation}
for $\sqrt{s}>1.7$ GeV.
The angular distribution of kaons in the $\pi N$ center of mass system
is taken as isotropic. Consequently, the Lorentz invariant double
differential kaon production cross section can be written as
\begin{equation}\label{pin}
{E\over p^2}{d^3\sigma (\sqrt s)\over dpd\Omega}=
\sigma_{K^+}(\sqrt{s}){E\over 4\pi p^2}\delta(p-q),
\end{equation}
where q is the kaon momentum in the $\pi N$ center of mass system
which is uniquely determined by the energy-momentum conservation.

With the above elementary kaon production cross sections, both the total
kaon production probability and the kaon Lorentz-invariant double differential
cross section in heavy-ion collisions can be calculated in the standard way.
In particular, the contribution to the kaon spectrum in the laboratory frame
from each elementary collision is calculated analytically according to
Eqs. (\ref{nn}) and (\ref{pin}) by performing successively two
Lorentz transformations.

\section{Results and discussion}
In the following we present and discuss results of our calculations on kaon
production probabilities and spectra. We concentrate on the four aspects
discussed in the introduction. It should be mentioned that the
momentum-independent, Skyrme-type mean field is used for all kinds of baryons.
Because kaon rescatterings have been found to affect mainly the
spectrum at large angles \cite{randrup81,ko}, and since we have in mind the
kaon
spectrum measured at $\theta_{lab}=44^{0}$ in the reaction of Au+Au at
$E_{beam}$/A=1.0 GeV, kaon rescatterings are negelected
in the present study.

\subsection{Sources of kaons in heavy-ion collisions}
An important question concerning kaon production in heavy-ion collisions
is where the kaons come from. For reactions at beam
energies around 1.0 GeV/nucleon, using the frozen-resonance approximation
it was shown previously that both the kaon yields and spectra
are dominated by $N\Delta, \Delta\Delta$ and $NN$ collisions sequentially.
To see how this conclusion may be affected by the finite
lifetime of baryon resonances we compare in parts (A) and (B) of Fig.\ 1
kaon production probabilities from various collision channels calculated using
the experimental, energy dependent resonance widths (A) with those calculated
using the frozen-resonance approximation (B).
The calculation was done for the reaction of Au+Au at a beam energy of
1.0 GeV/nucleon and an impact parameter of 1.0 fm. The soft nuclear
equation of state corresponding to an incompressibility $K=200 $ MeV is used.
In both cases the Lorentz transformation is used to boost the initial momentum
distributions of the two colliding nuclei into their center of mass frame.

There are several interesting points to be noticed. First,
we notice that the total kaon production probabilities in the
two calculations are comparable ($P_{A}=0.0653$ and $P_{B}=0.0686$).
The main difference lies in the relative contributions to the kaon yield and
the shapes of the kaon spectrum. The kaon production probabilities from
both the $N\Delta$ and the $\Delta\Delta$ collision channel are reduced
by about 50\% when finite lifetimes are used for the resonances. The $\pi N$
contribution is about 25\% of the total kaon yield. This is in agreement with
previous studies\cite{cugnon,batko90,xiong90}. However, it is seen that the
$\pi N$ contribution is comparable to contributions of most baryon-baryon
collision channels and needs to be included. As we will discuss later,
the $\pi N$ contribution to the kaon spectrum is significantly different
from that of baryon-baryon collisions due to the rather different
energy dependence of the elementary kaon production cross sections and
the reaction kinematics.

Second, the kaon production probabilities from the $N^{*}$ involved
collisions is increased when finite lifetimes are used for
baryon resonances. This is because the $N^{*}$ resonance is mainly excited
through $\pi N$ collisions, while the excitation through NN collisions is
negligibly small at beam energies around 1.0 GeV/nucleon.

Third, we notice that the contribution from $\Delta\Delta$ collisions
is small compared to that from NN collisions even in the case of using the
frozen-resonance approximation. This is in contrast to some of the previous
calculations where the contribution from $\Delta\Delta$ dominates over
that from NN collisions \cite{huang,aichelin}. The origin of this
difference remains to be explored. One of the possible reasons is the
use of the non-relativistic momentum transformation in some of
the previous studies.

In view of the fact that the non-relativistic transformation was used in
some of the previous calculations to boost the initial Fermi momentum
distributions of the two colliding nuclei into their center of mass frame,
we study now effects of the momentum transformation.
The importance of the nuclear Fermi motion in subthreshold particle
production is well konwn. The free threshold beam energy for kaon
production is 1.58 GeV; it is reduced to about 0.8 GeV after taking
into account the Fermi motion \cite{aichelin85}. In transport models or
quantum molecular dynamics models, nuclear reactions are simulated
in the nucleus-nucleus center of mass frame. The boosting of the two
Fermi spheres has to be treated with care. At relativistic energies,
the boost must be done with the Lorentz transformation. For example,
for the reaction of Au+Au at $E_{beam}/A=1.0 $ GeV, the Lorentz $\gamma$ factor
is 2.1, the non-relativistic transformation supresses the kinetic
energy of each nucleon in the nucleus-nucleus center of mass frame
by about 35 MeV on the average.

To be more quantitative, we presnt in parts (C) and (D) of Fig.\ 1
the kaon production probabilities calculated using the
non-relativistic momentum transformation. Results in (C) are calculated
with finite lifetimes of baryon resonances
and that in (D) are calculated with the infinite lifetime.
The total kaon production probability in case C is $P_{C}=0.0453$
and that in case D is $P_{D}=0.0406$.
Comparing them with results shown in (A) and (B), it is seen that the total
kaon
yields decrease by about 40\% when the non-relativistic momentum transformation
is used. In particular, the kaon yield from NN collisions calculated with
the relativistic transformation is about twice that calculated using
the non-relativistic transformation. We notice that the change of
the kaon yield due to the momentum transformation is comparable to
that caused by changing the nuclear equation of state.
This can be seen from Fig.\ 2 where the kaon production probability
calculated using an incompressibility $K=380$ MeV and the
relativistic momentum transformation is shown.
In this case the total kaon production probability is $P= 0.0360$.
It is seen that the total kaon yield is reduced by about 45\% by changing
the incompressibility from $K=200 MeV$ to 380 MeV.
It is therefore clear that the seemingly
numerical treatment of the initial momentum transformation
has an important physical consequence. Kaon yields
calculated previously using the non-relativistic transformation for
the reaction of Au+Au at $E_{beam}/A=1.0$ GeV should be increased
correspondingly.

Comparing results calculated using the finite and infinite lifetimes for baryon
resonances, it is seen that the total kaon yield changes only by about
5\% to 12\% although the relative contributions from the various collision
channels change significantly. The shape of the kaon spectrum, however, varies
significantly. The kaon spectra calculated in the laboratory frame
are shown in Fig.\ 3 for the cases A, B, C and D.
It is noticed that the frozen-resonance approximation underestimates
the high momentum part of the kaon spectra. This is easily understandable.
On the average the center of mass of a $\pi N$ pair moves faster than that of
a baryon-baryon pair due to the lighter mass. The kaon
from the low threshold ($\sqrt{s}=1.61$ GeV) but two-body
final state $\pi N\rightarrow \Lambda K$ process has the same average
energy as the kaon from the high threshold ($\sqrt{s}=2.55$ GeV) but
three-body final state $BB\rightarrow B Y K$ process. Therefore,
the kaon from $\pi N$ collisions can have higher laboratory momenta
than the kaon from baryon-baryon collisions.

\subsection{Centrality dependence of kaon production}
In this section we concentrate on the centrality dependence of
the total kaon production probabilities. In Fig.\ 4 we present kaon
production probabilities from various channels as a function
of impact parameter for both the soft (left) and the stiff (right)
equations of state. The rapid decrease of the kaon production probability
with the increasing impact parameter is obvious. By changing from the soft
nuclear equation of state to the stiff one, the total kaon production
probability changes by about 40\% to 50\% in the whole impact
parameter range. The most significant change
happens to the relative contributions from the $N\Delta$ and $NN$ collisions.
With the soft equation of state, the contribution from $N\Delta$ collisions
is slightly higher than that from $NN$ collisions in central collisions.
The two contributions are comparable in peripheral collisions.
While with the stiff equation of state, the contribution from $NN$ collisions
dominates over the contribution from $N\Delta$ collisions. The contribution
to kaons from the $NN$ collisions is reduced by about 27\%, while that from
the $N\Delta$ collisions is reduced by about 44\% comparing to the calculation
with the soft equation of state. The overall reduction of the kaon yields
is due to the lower compression reached in the reaction
and the smaller energy available for particle production or excitation of
nucleons with the stiff equation of state.
Furthermore, since the production of kaons from $N\Delta$ collisions requires
two successive interactions ($NN\rightarrow N\Delta$ and
$N\Delta\rightarrow NYK$), kaons from the $N\Delta$ collisions are
more sensitive to the change of the nuclear equation of state.

\subsection{Effects of $N^{*}(1440)$ resonances on kaon production}
Experimental data on particle production accumulated
at several laboratories during the last decade
indicate that a gradual transition to resonance matter
occurs in the participant region of heavy-ion collisions at beam energies
of 1 to 2 GeV/nucleon \cite{metag}.
To study in detail properties of this new form of matter it is
important to determine its baryonic composition, i.e., the relative populations
of nucleons and various baryon resonances. In the 1-2 GeV/nucleon
beam energy range, the important baryonic resonances are $\Delta(1232)$,
$N^{*}(1440)$, $N^{*}(1520)$ and $N^{*}(1535)$. The excitation of the
$\Delta(1232)$ and $N^{*}(1535)$ resonances have been studied extensively
through the production of pions and etas respectively \cite{mosel,berg}.
The study on the $N^{*}(1440)$ resonance has been rather rare although it is
the second most important baryon resonance to be excited in the energy range
considered here. Here we discuss briefly effects of the $N^{*}(1440)$
resonances on kaons, more detailed study of the $N^{*}$ resonance together
with its effects on subthreshold antikaon and antiproton production will be
published elsewhere\cite{li94}.

First, to evaluate the importance of the excitation of the
$N^{*}(1440)$ resonance we show in Fig.\ 5
the $N^{*}(1440)$ multiplicity during the Au+Au collisions at
beam energies of 1 to 2 GeV/nucleon and impact parameters
between 1 and 9 fm. It is seen that the population of the $N^{*}(1440)$
resonance strongly depends on the impact parameter and the beam energy.
In central collisions the maximum number of $N^{*}(1440)$ resonances increases
from 7 to 17 as the beam energy increases from 1 to 2 GeV/nucleon.
For comparison, it should be mentioned
that the maximum number of the $\Delta(1232)$ resonance increases from 52 to 89
in the same collisions and therefore the ratio of the populations of the
$N^{*}$ and the $\Delta$ increases from 14\% to 19\% as the beam energy
increases from 1 to 2 GeV/nucleon.

In Fig.\ 6, we show the total and the $N^{*}$
induced kaon production
probabilities in Au+Au collisions at
beam energies between 1 and 2 GeV/nucleon and impact parameters
between 1 and 9 fm. The total kaon production probabilities
are shown with the solid lines while the contributions from the $N^{*}$
involved
collisions are shown with the dashed lines. It is seen that,
the $N^{*}$ involved collisions contribute only about 11\% in
the whole energy and impact parameter ranges. This result is not
surprising since baryon-baryon collisions at
center of mass energies around the kaon production threshold of
$\sqrt{s}=2.55$ GeV are dominated by the production and
absorption of $\Delta(1232)$ resonances. Effects of the $N^{*}$ resonance on
the kaon spectrum will be discussed later.

\subsection{Pion spectra}
{}From what we have discussed above, it is clear that the production of
kaons is virtually determined by the $\pi,N,\Delta,N^{*}$ dynamics of
the reaction. Since kaons are calculated perturbatively during the reaction,
to extract reliably the interesting physics from kaon observables it is a
prerequisite that the reaction dynamics is correctly described.
This can be done by studying pion observables, such as pion spectra.
For this purpose we discuss in this section the pion spectra.
Fortunately, the kaoS collaboration
have measured both the kaon and pion spectra in the same
experiments \cite{grosse,senger}.

It is well known that pion spectra are rather insensitive to the
nuclear equation of state since the total pion multiplicity calculated
with different equations of state differs only by about 15\%. Here we
show pion spectra calculated with the soft equation of state.
In Fig.\ 7 and Fig.\ 8 we compare the calculated $\pi^{+}$ and $\pi^{-}$
spectrum with the experimental data measured at
$\theta_{lab}=44^{0}$ in the reaction of Au+Au at $E_{beam}/A=1.0 $ GeV.
The calculations are shown by open squares with error bars.
The available experimental data from the KaoS collaboration
are given by the solid circles \cite{grosse,senger,muntz}. It is seen that
the data can be very well reproduced. For completeness and future
comparison we present the model prediction of $\pi^{0}$ spectrum in Fig.\ 9.
The comparison between the model calculation and the experimental data on
pion spectra indicates that the complicated reaction dynamics is well
described in the model.

\subsection{Kaon spectra}
We now turn to our study on kaon spectra by comparing
the kaon spectra from different collision channels and performing a
comparison between the model calculation and the experimental data.

Kaon spectra from the seven collision channels
($N\Delta, NN, \Delta\Delta, NN^{*}, N^{*}N^{*}, \Delta N^{*}$ and $\pi N$)
calculated at $\theta_{lab}=44^{0}$ in the reaction of Au+Au at
$E_{beam}/A=1.0$ GeV are displayed in Fig.\ 10. The right window
shows results calculated using the soft nuclear equation of state
while the left window shows that calculated
using the stiff equation of state. It is interesting to note that
in the low momentum part ($P_{lab}\leq 0.5$ Gev/c) $N\Delta$ and
$NN$ collisions dominate, while in the high momentum part $\pi N$ and
$NN^{*}$ collisions dominate. The fact that $\pi N$
collisions are the most important source for high
momentum kaons is mainly due to the reaction kinematics as we have
explained early. It should also be mentioned that the kaon spectra
from $\pi N$ collisions have larger statistical fluctuations due to the
resonance like $\pi N\rightarrow \Lambda K$ cross section.
It is also seen that the overall effect of the $N^{*}$ resonace on
kaon spectra is small. However, it is interesting to mention that
the contribution from $NN^{*}$ collisions is much larger than that from
$\Delta\Delta$ collisions, especially in the high energy part of the spectra.
This is mainly due to the large $\rho_{N}/\rho_{\Delta}$ ratio and the higher
mass of the $N^{*}$ resonances.

We now perform a comparison between the model calculation and the experimental
data in Fig.\ 11. The experimental kaon spectrum measured at
$\theta_{lab}=44^{0}$ in the reaction of Au+Au at $E_{beam}/A=1.0$ GeV
is ploted with solid circles. The open circles and squares are
calculated with the soft and stiff equations
of state respectively. It is seen that the experimental data can be well
reproduced with the soft equation of state. This finding is in agreement
with that of ref.\cite{aichelin} although the sources of kaons are quite
different. The fact that both the pion and the kaon spectrum can be well
reproduced simultaneously indicates that the production of kaons is
mainly determined by the $\pi,N,\Delta,N^{*}$ dynamics of
the reaction and it has been well accounted for in the model, in addition
the elementary cross sections used for kaon production are reasonable.

It should, however, be stressed that
the agreement between the calculation and the data cannot be used
to extract the nuclear equation of state since the rescattering of kaons
although its effect is samll at this angle, and moreover, neither the momentum
dependent part of the mean field nor the in-medium effects on kaons and
nucleons are taken into account in the present model calculations. It has
been shown that the momentum dependent interaction reduces the kaon yield
by a factor of 2 to 3 \cite{aichelin}. On the other hand, more consistant RBUU
calculations includeing both the momentum-dependent mean field and the
medium effects have shown that the medium effects on kaons results in an
enhancement of kaon yields by about a factor of 2 to 3 \cite{ko}. It would
therefore be interesting to study in the future whether the model used
here can still reproduce the data with both the momentum dependent
mean field and the in-medium effects. However, due to the still rather
model dependent prescription of the momentum dependent mean field
and the in-medium effects it has been difficult to compare
results from different groups and to extract reliably the interesting
physics from the kaon data. It seems therefore plausible to first
compare calculations with the least and widely accepted model approximations.
The present model calculations and associated
discussions on several basic aspects of kaon production are useful
in this respect.

\section{Summary}
Using a hadronic transport model we have studied several aspects of
near-threshold kaon production in relativistic heavy-ion collisions.
We showed that the finite lifetime of baryon resonances affects significantly
the shape of the kaon spectra. $\pi N\rightarrow \Lambda K$
channel is found to be the most important source for high energy kaons.
$N^{*}(1440)$ resonances contribute about 10\% to the total kaon yield
and have a minor effect on the kaon spectra. The initial Fermi momentum
transformation has an effect on kaon yields and spectra as large as that of the
nuclear equation of state. With the present model, both the pion and the
kaon spectra from the reaction of Au+Au at $E_{beam}/A=1.0$ GeV can
be well reproduced simultaneously with the soft nuclear equation of state.

The author would like to thank Prof. C.M. Ko and Dr. G.Q. Li for
stimulating and fruitful discussions during the course of this work
and their critical reading of the manuscript. He is also grateful to Drs.
H. Oeschler, Ch. M\"untz and A. Wagner for helpful discussions.

\newpage
\section*{Figure Captions}

\begin{description}

\item{\bf Fig.\ 1}\ \ \ Time evolution of kaon production probabilities
in the reaction of Au+Au at $E_{beam}/A=1.0$ GeV and b=1.0 fm using
the soft equation of state.
(A) Using finite lifetime for resonances and the relativistic
momentum transformation
(B) Using infinite lifetime for resonances and the relativistic
momentum transformation
\item{\bf Fig.\ 1 continued}\ \ \
(C) Using finite lifetime for resonances and the non-relativistic
momentum transformation
(D) Using infinite lifetime for resonances and the non-relativistic
momentum transformation

\item{\bf Fig.\ 2}\ \ \
Time evolution of kaon production probabilities for the Au+Au reaction
using the stiff equation of state.

\item{\bf Fig.\ 3}\ \ \
Kaon momentum spectra corresponding to the cases A, B, C and D in Fig.\ 1.

\item{\bf Fig.\ 4}\ \ \
Impact parameter dependence of kaon production probabilities for the
reaction of Au+Au at $E_{beam}/A=1.0$ GeV for both the soft and stiff equation
of states.

\item{\bf Fig.\ 5}\ \ \
Time evolution of the $N^{*}(1440)$ populations in the Au+Au collisions.
(left) Impact parameter dependence. (right) Beam energy dependence.

\item{\bf Fig.\ 6}\ \ \
The total (solid) and $N^{*}(1440)$ induced (dashed) kaon production
probabilities in the Au+Au collisions. (left) Impact parameter dependence.
(right) Beam energy dependence.

\item{\bf Fig.\ 7}\ \ \ $\pi^{+}$ spectra for the reaction of Au+Au
at $E_{beam}/A=1.0$ GeV.

\item{\bf Fig.\ 8}\ \ \ $\pi^{-}$ spectra for the reaction of Au+Au
at $E_{beam}/A=1.0$ GeV.

\item{\bf Fig.\ 9}\ \ \ $\pi^{0}$ spectra for the reaction of Au+Au at
$E_{beam}
/A=1.0$ GeV.

\item{\bf Fig.\ 10}\ \ \ Kaon spectra from 7 different collision channels
in the reaction of Au+Au at $E_{beam}/A=1.0$ GeV.

\item{\bf Fig.\ 11}\ \ \ Comparison between the model calculations and the
experimental data on kaon spectra from the reaction of Au+Au at $E_{beam}/A$
=1.0 GeV.

\end{description}

\end{document}